\documentclass[twocolumn,prl,superscriptaddress]{revtex4-1}
\usepackage{palatino}
\usepackage[latin1]{inputenc}
\usepackage{epsf}
\usepackage{amsmath,amssymb}
\usepackage{latexsym}
\usepackage{calc}
\usepackage{xcolor}
\usepackage{graphicx}
\usepackage{hyperref}
\newcommand{\ben}{\begin{equation}}
\newcommand{\een}{\end{equation}}
\newcommand{\bea}{\begin{eqnarray}}
\newcommand{\eea}{\end{eqnarray}}

\def\ket#1{\vert#1\rangle}

\def\sss{\scriptscriptstyle\rm}

\def\1s{_{1,\sss S}}
\def\2s{_{2,\sss S}}
\def\x{_{\sss X}}

\def\s{_{\sss S}}
\def\xc{_{\sss XC}}

\def\H{_{\sss H}}

\def\br{{\bf r}}
\def\bR{{\bf R}}

\def\bR{{\bf R}}
\def\bd{{\bf d}}

\def\dulr{{\underline{\underline{\bf r}}}}
\def\dulR{{\underline{\underline{\bf R}}}}

\begin{document}
\title{Minimizing the time-dependent density functional error in Ehrenfest dynamics}
\author{Lionel Lacombe}
\affiliation{Department of Physics, Rutgers University, Newark 07102, New Jersey USA}
\author{Neepa T. Maitra}
\affiliation{Department of Physics, Rutgers University, Newark 07102, New Jersey USA}
\email{liolacombe@gmail.com, neepa.maitra@rutgers.edu}
\date{\today}

\begin{abstract}
Simulating electron-ion dynamics using time-dependent density functional theory within an Ehrenfest dynamics scheme can be done in two ways that are in principle exact and identical: propagating time-dependent electronic Kohn-Sham equations  or propagating electronic coefficients on surfaces obtained from linear-response. We show here  that using an approximate functional leads to qualitatively different  dynamics in the two approaches. We argue that the latter is more accurate because the functionals are evaluated on domains close to the ground-state where current approximations perform better. We demonstrate this on an exactly-solvable model of charge-transfer, and discuss implications for time-resolved spectroscopy. 
\end{abstract}

\maketitle
Coupled electron-nuclear dynamics lies at the heart of several topical phenomena, including photovoltaic design, photocatalysis, or the  laser control of chemical reactions. To accurately simulate these processes computationally, adequate accounting of both electron-nuclear correlation as well as electron-electron interactions is needed in practical dynamics schemes. 
From the  purely electronic structure side, time-dependent density functional theory (TDDFT)~\cite{RG84,TDDFTbook2012,Carstenbook,M16} is a practical choice when more than a few atoms are involved: one solves a system of non-interacting electrons, the Kohn-Sham system, in which many-body interaction effects are ``hidden" in the exchange-correlation (xc) functional.
TDDFT has yielded useful agreement with experiment in an impressive array of cases, including when coupled to nuclear motion~\cite{RFSR13,T15,MPBLZF17}. 
From the electron-nuclear coupling side, the mixed quantum-classical methods of Ehrenfest dynamics and trajectory surface-hopping are currently most widely-used. While Ehrenfest dynamics is unable to capture effects like wavepacket splitting, it has several desirable features that make it more attractive to use in many cases, including that it is cheaper and faster since it is does not rely on a stochastic algorithm, which enables calculations on  large systems (see e.g. Ref~\cite{DAGBSC17} for a computation involving 59 400 number of electrons). Further it is derivable from first-principles and so does not have to make somewhat {\it ad hoc} choices such as velocity-rescaling that surface-hopping does. 

TDDFT is an ideal partner for Ehrenfest dynamics since the coupling of the electronic system to the nuclear one is directly via the electronic density, so no additional ``observable functionals"~\cite{M16} need to be extracted from the Kohn-Sham system. It is available in several widely-used codes, such as Octopus\cite{octopus_2020}, NWChem~\cite{nwchem_2020}, Sharc\cite{sharc_2018}, Newton-X\cite{newton-x_2014}, Quantum Espresso\cite{quantum_espresso_2020}, Salmon\cite{salmon_2019}, Siesta\cite{siesta_2020}, or CPMD~\cite{CP85}, where it is implemented in one of two distinct ways: one involving propagation of time-dependent Kohn-Sham orbitals, and the other propagation of coefficients involving energies and couplings obtained from TDDFT linear response. 
When using the same approximate functional, one would hope to get  similar results from the different codes, all else (e.g. basis sets, convergence thresholds) being the same. However,  we show here that the two distinct implementations lead to qualitative differences in the resulting dynamics. Fundamentally, this is because the domains that the xc functional is evaluated  on are fundamentally different. One implementation probes the functional on the fully non-equilibrium time-dependent density, and the true and KS wavefunctions underlying the density at any time are typically not ground-states. 
The other implementation needs to evaluate it and its functional derivative merely on a density whose underlying state is a ground-state at the instantaneous nuclear configuration. Because approximate functionals tend to be more accurate in the latter case, the resulting Ehrenfest dynamics is also much more accurate when the implementation is done in this way. We demonstrate this explicitly on a model system of photo-excited charge-transfer where a comparison with exact dynamics can be made. 
We discuss the significance of these results for time-resolved spectroscopy.

In Ehrenfest dynamics~\cite{T98}, the nuclei are described by an ensemble of classical trajectories evolving on an averaged potential energy surface (PES) determined by the electrons:
\ben
M_\nu\ddot\bR^{(I)}_\nu = -\langle\Psi^{(I)} \vert\nabla_\nu H_{\rm BO} \vert \Psi^{(I)} \rangle 
\een
where $\Psi^{(I)}$ is the electronic wavefunction following the electronic TDSE, $H_{\rm BO}(R^{(I)}(t))\Psi^{(I)} = i \partial_t \Psi^{(I)}$,  with the instantaneous position of trajectory $I$ appearing in the electron-nuclear coupling term in the Born-Oppenheimer (BO) Hamiltonian.
Noting that the nuclear gradient operates only on the electron-nuclear interaction potential, $V_{en} = \sum_n^{N_e}\sum_\nu^{N_n} v_{en}(\vert \br_n - \bR_\nu\vert)$, which is a multiplicative one-body operator from the standpoint of the electrons, we may re-write this as
\ben
M_\nu\ddot\bR_\nu = - \int d^3r n(\br,t) \nabla_\nu V_{en}(\dulr,\dulR) - \nabla_\nu V_{nn}(\dulR)
\label{eq:RddotRT}
\een
where $n(\br,t) = N\sum_\sigma \int d^3 r_2... d^2 r_N \vert\Psi(\br, \br_2...\br_N)\vert^2$ is the one-body electron density and $\dulr, \dulR$ denote coordinates of all the electrons and nuclei respectively. The problem then appears perfectly suited to work with TDDFT, where, instead of having to solve a many-body interacting TDSE, the $n(\br,t)$ is produced by evolving one-body TDKS equations~\cite{RG84,TDDFTbook2012,Carstenbook}
\ben
\left( -\nabla^2/2+ v\s(\br,t)\right)\phi_k(\br,t) = i \partial_t \phi_k(\br,t),
\label{eq:TDKS}
\een
with $n(\br,t) = \sum_{k \in occ.}\vert\phi_k(\br,t)\vert^2$ and 
\ben
v\s(\br,t) = v_{en}(\br,\dulR^{(I)}(t)) + v\H[n](\br,t) + v\xc[n; \Psi_0,\Phi_0](\br,t).
\label{eq:vs}
\een
Here $v\H[n](\br,t) = \int \frac{n(\br',t)}{\vert \br - \br'\vert} d^3r'$ is the Hartree potential, and $v\xc$ is the xc potential, in principle a functional of the initial interacting state $\Psi_0$, the initial choice of KS orbitals $\Phi_0$, and the history of the density. 
Armed with an approximation for $v\xc$, Eqs.~(\ref{eq:RddotRT}) and (\ref{eq:vs}) are solved together, with the electronic and nuclear dynamics coupled through the electronic density and nuclear position, and we refer to this approach as ``real-time (RT) Ehrenfest".

It is equivalent to instead expand $\ket{\Psi(t)}$  in terms of the exact interacting BO eigenstates and evolve the coefficients: writing $\ket{\Psi^{(I)}(t)} = \sum_j C_j(t)\ket{\Psi^{\rm BO}_{\dulR^{(I)}(t), j}}$, where $ H_{\rm BO}\Psi^{\rm BO}_{\dulR,j} = E_j(\dulR)\Psi^{\rm BO}_{\dulR^{(I)}, j}
$ we have
\ben
\dot{C}_j^{(I)}(t) = -i E_j^{(I)}(t)C_j^{(I)}(t)  - \sum_\nu\sum_k{\bd}_{jk,\nu}^{(I)}\cdot \dot\bR_\nu^{(I)} C_k^{(I)}
\label{eq:coeffs}
\een 
using the short-hand $E_j^{(I)}(t) = E_j(\dulR^{(I)}(t))$, and the non-adiabatic couplings $\bd_{jk,\nu} = \langle \Psi_j^{\rm BO} \vert \nabla_\nu\Psi_k^{\rm BO}\rangle$. 
In terms of these coefficients, the force on the nuclei takes the form of the weighted average 
\ben
\begin{split}
M_\nu\ddot\bR^{(I)}_\nu 
=
-&\sum_j \vert C_j^{(I)}(t)\vert^2 \nabla_\nu E_j^{(I)}(t)
\\
-&\sum_{jk} C_j^{(I)*}(t)C_k^{(I)}(t) (E_k^{(I)}(t)-E_j^{(I)}(t)){\bd}_{jk,\nu}^{(I)}
\end{split}
\label{eq:RddotLR}
\een
Eqs.~(\ref{eq:coeffs})--(\ref{eq:RddotLR}) present an alternative implementation of Ehrenfest dynamics, and can also be used in conjunction with TDDFT, where linear response gives the energies $E_j(R)$ and couplings  $\bd_{jk,\nu}$ between ground and excited states, with quadratic response giving the couplings between excited states~\cite{PRF18,NCAJG17}. We denote the approach Eqs.~(\ref{eq:coeffs})--(\ref{eq:RddotLR}) as ``linear-response (LR) Ehrenfest". 

In theory, the RT-Ehrenfest scheme Eqs.~(\ref{eq:RddotRT})--(\ref{eq:vs}) and the LR-Ehrenfest scheme Eqs.~(\ref{eq:coeffs})--(\ref{eq:RddotLR}) are entirely equivalent, and, if the exact functionals were known and used, would yield identical results for both electronic and nuclear observables,  corresponding to running Ehrenfest dynamics with an exact electronic structure method. 

However in practise, there is a fundamental difference due to the domains on which the xc functionals in the two approaches is evaluated. 
While LR-Ehrenfest requires functionals for the xc energy and xc kernel evaluated on the density obtained from a ground-state, RT-Ehrenfest requires the functional for the xc potential to be evaluated on the fully non-equilibrium time-dependent density where the underlying states of the true system and KS system are far from any ground state.  Since functional approximations in use today are predominantly adiabatic, i.e. built from ground-state ones, they tend to perform much better in the LR regime than in non-perturbative situations, and so we expect that the LR-Ehrenfest scheme  will give more reliable and accurate results than the RT-Ehrenfest scheme. 

This situation is not dissimilar to TDDFT simulations of the purely electronic scattering problem~\cite{LSWM18} where a time-resolved calculation of electron-molecule scattering dynamics probes the xc functional in a fully non-linear regime, giving far poorer scattering probabilities than  a formulation extracting these same probabilities from  linear response. Although  the response-based formulation is only valid in the elastic case and cannot provide a time-resolved picture, the results in the elastic case are better than in the real-time calculation. 

Returning to the electron-ion dynamics, we note that the difference between the two schemes persists even when there is negligible non-adiabatic coupling such that Ehrenfest reduces to BO. That is, even in cases where dynamics is such that the nuclear trajectories evolve on a single excited BO PES, the LR-BO and RT-BO methods will give different answers when approximate functionals are used. They will only agree when the dynamics takes place on purely the ground-state PES.

\begin{figure}
\includegraphics[width=0.5\textwidth]{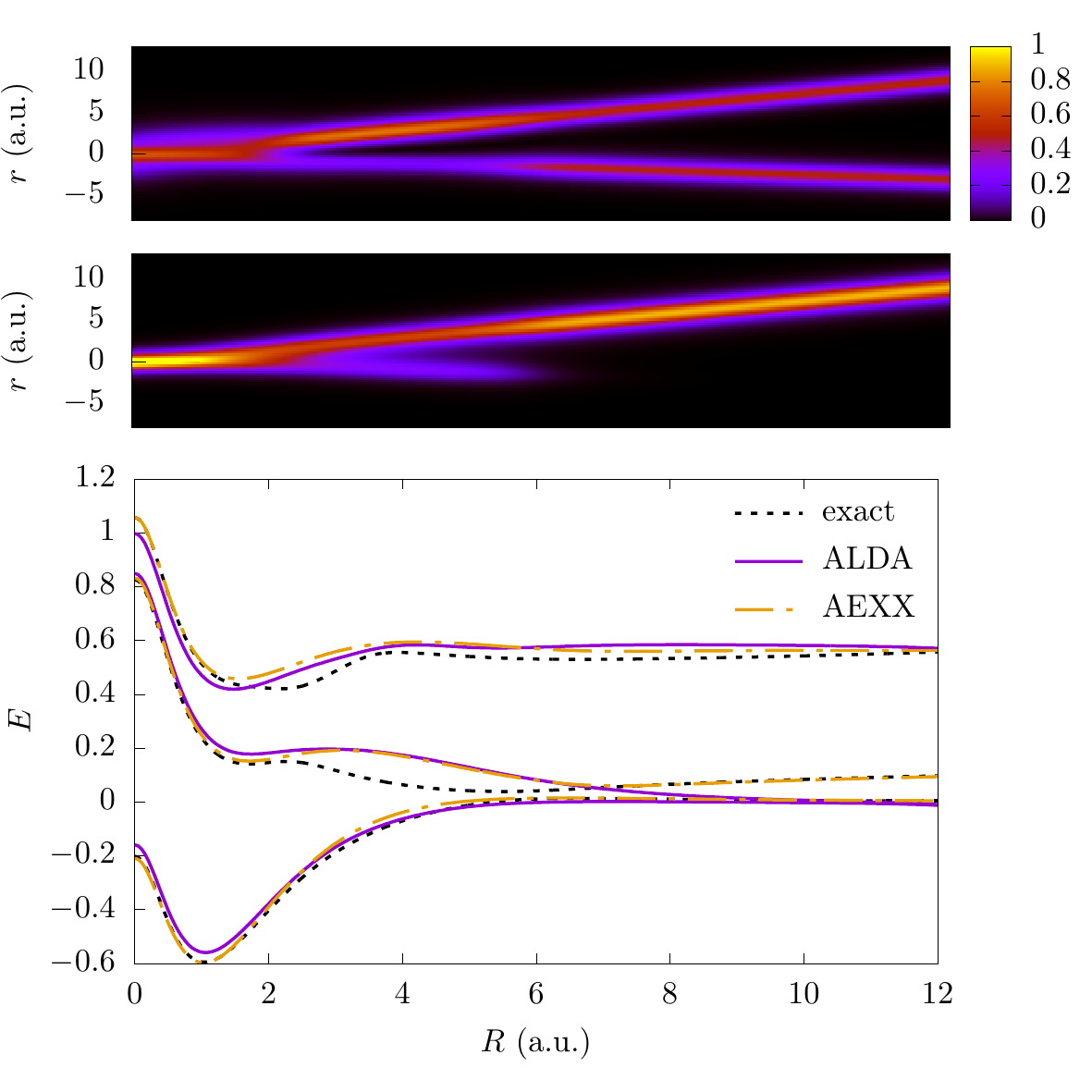}
 \caption{
   Lower panel: Lowest three singlet BO PESs, exact, linear-response ALDA and AEXX BO PESs as indicated in the legend. Upper panel, top to bottom: heat map of exact $\vert \Psi_{R,1}^{\rm BO}(r) \vert^2$ and $\vert \Psi_{R,0}^{\rm BO}(r) \vert^2$.
 }
 \label{fig:surfaces}
\end{figure}

We use an exactly-solvable model system simulating a photo-excited ion-driven charge-transfer event to explicitly demonstrate the problem. Consider a one-dimensional molecule consisting of two soft-Coulomb interacting electrons, an ion with net charge $Z=2$ and mass $m_1 = 2 m_p$ and an atom of zero net charge and mass $m_2 = 6 m_p$, with $m_p = 1836.1528$ a.u. being the mass of the proton. The ion may be thought of as a bare nucleus, and its interaction with the electrons is taken as soft-Coulomb. The atom may be thought of as a nucleus surrounded by a frozen cloud of electrons, such that it presents a short-ranged screened soft-Coulomb potential to the electrons. The  ion-atom is also taken as screened soft-Coulomb; details are given in the Supplementary Material. 
In the dissociation limit, the ground-state captures both electrons in the soft-Coulomb well, while the first-excited state  represents a charge-transfer state with one-electron in the ionic well and the other in the atomic well.  Fig.~\ref{fig:surfaces} shows these two BO states, as well as the lowest three BO PES as a function of ion-atom separation $R$. We show also the surfaces obtained from DFT and LR TDDFT  using two contrasting functional approximations: adiabatic local density approximation (ALDA)\cite{HFCVMTR11,CSS06} which has local dependence on the density in time and in space, and adiabatic exact exchange (AEXX) which has a non-local dependence on the density in space, but still local in time. In this two-electron case, $v\x[n](\br,t) = -v\H[n](\br,t)/2$. 
We observe that they both approximate the ground-state energy quite well for all $R$. For large $R$, AEXX captures the asymptotic $I_D - A_A -1/R$ tail of the exact charge-transfer state well, while ALDA collapses to zero, as expected~\cite{M17}. The ALDA ground and first excited orbitals become degenerate and delocalized over the molecule. At intermediate distances, we see that ALDA and AEXX excited BO PES are very similar to each other,  with a more pronounced shape resonance than the exact. 

The eigenvectors of the TDDFT linear response matrix indicate that the first excitation is so significantly dominated by the lowest KS excitation, that a small matrix approximation~\cite{PGG96,GPG00,TDDFTbook2012} yields practically the same curves for both functional approximations; higher excited states mix in negligibly.   The density of the excited state is then very well-approximated by the density of the KS excited determinant, which will ease our task in finding the electronic dipole and densities from LR-Ehrenfest calculations. 

\begin{figure}
\includegraphics[width=1.0\columnwidth]{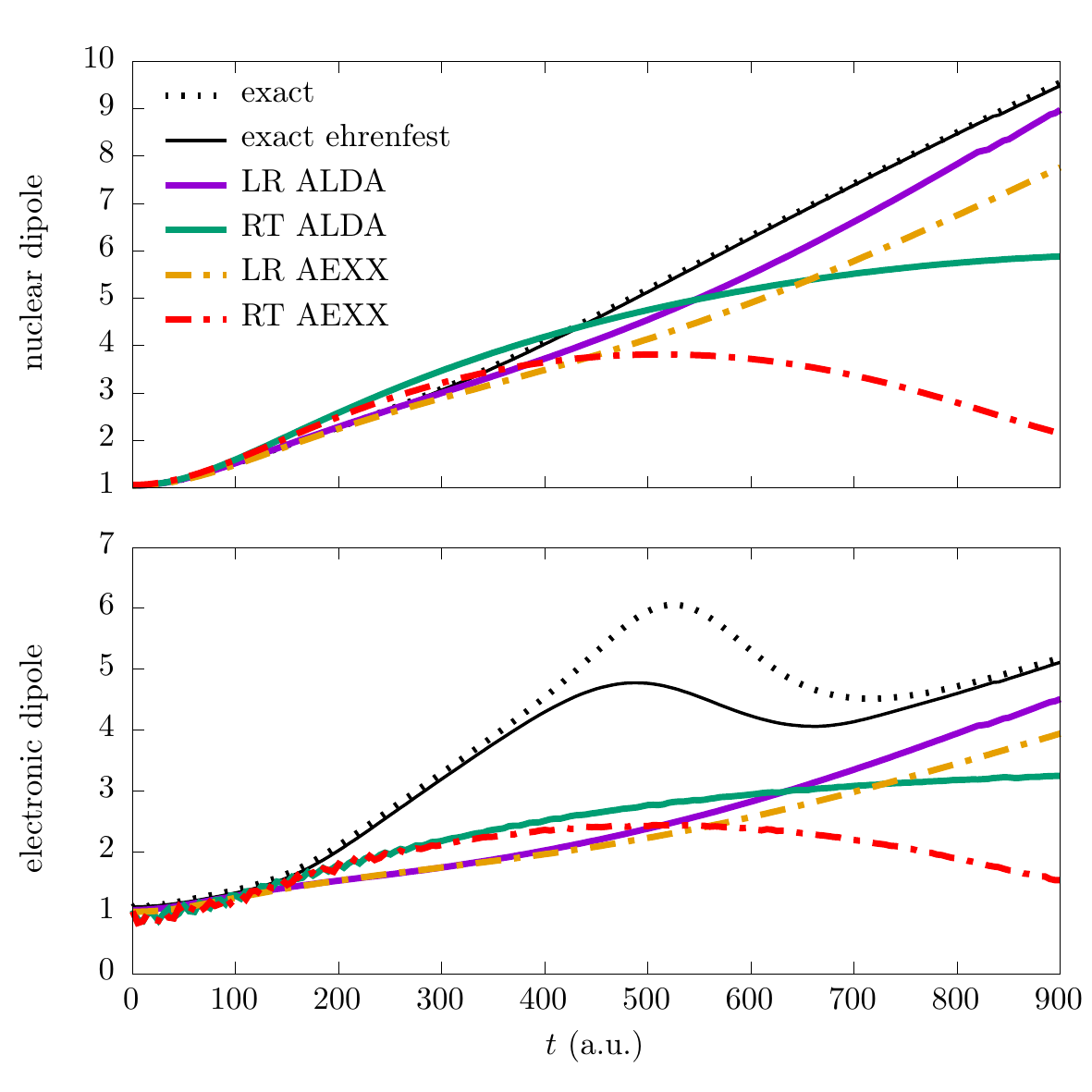}
 \caption{
   Time-dependent nuclear (upper panel) and electronic (lower panel) dipoles: exact, exact Ehrenfest, and the TDDFT LR-Ehrenfest and RT-Ehrenfest calculations: LR ALDA, RT ALDA, LR AEXX, RT AEXX as indicated in the legend.
 }
 \label{fig:dips}
\end{figure}

We now begin the dynamics from an initial photo-excitation that lifts the ground-state nuclear wavefunction of the ground BO state to the first excited electronic state. Due to the slope of the first excited surface, the nuclear wavepacket moves towards larger $R$, but slowly enough such that as it passes  the avoided crossing region around $R \approx 5.8$a.u. it remains largely in the excited state. Only a small fraction of the wavepacket transfers to the ground-state surface: in the exact dynamics, the maximum population of the ground BO state is $0.166$ which then largely transfers back, leaving only $0.03$ ground-state population  at long times. In an Ehrenfest calculation using the exact surfaces, even less, $0.04$, transfers to the lower surface near the avoided crossing, while at long times only $0.03$ remains there. 
Thus Ehrenfest dynamics practically reduces to BO dynamics on the first excited state. The exact nuclear dipole moment shown in Figure~\ref{fig:dips} agrees closely with that of a 100-trajectory RT Ehrenfest calculation using forces derived from the exact first BO surface, denoted ``exact Ehrenfest". The lower panel shows the electronic dipoles are close except as the avoided crossing region is passed, where the exact density has a larger ground-state component than the Ehrenfest one, as noted above. The difference is less remarkable in the electronic densities, see Fig.~\ref{fig:snaps} and a movie in the Supplementary Material. 

Turning now to the TDDFT approximations shown in Fig~\ref{fig:dips}, an immediate observation is the qualitatively incorrect behavior of the RT calculations for both the nuclear and electronic dipoles: after their initial rise, these curve downwards, instead of continuing to move to larger values; ultimately neither dissociate nor transfer the electron.
The RT ALDA is a better approximation than the RT AEXX; 
this is also true for the electronic density shown in Fig.~\ref{fig:snaps} (see also the movie in the SI, which also shows the KS potentials for the two  cases). The LR TDDFT calculations perform qualitatively much better, even if they  undershoot both the nuclear and electronic dipoles, largely due to the wrong shape of the linear response curve in the intermediate region $R\approx 2 - 7$a.u. (Fig.~\ref{fig:surfaces}). Around $R = 5.8$ a.u. the density of the exact first-excited state switches from being mainly localized on the ion to the 50:50 charge-transfer density, as evident in Fig.~\ref{fig:surfaces} and reflected in the step in the exact and exact-Ehrenfest electronic dipole in Fig.~\ref{fig:dips}. On the other hand the ALDA and AEXX excited state densities do not have this feature, showing instead a smoother charge-transfer and corresponding dipole. 
It appears the  LR ALDA dipoles and densities are better than LR AEXX (except at  large times where the LR ALDA nuclei move too fast due to the lack of the $-1/R$ tail); the LR ALDA benefits from a partial cancellation of errors in that effect of the underestimated slope at intermediate time is compensated by the lack of the $-1/R$ at later times.

 \begin{figure}
\includegraphics[width=1.0\columnwidth]{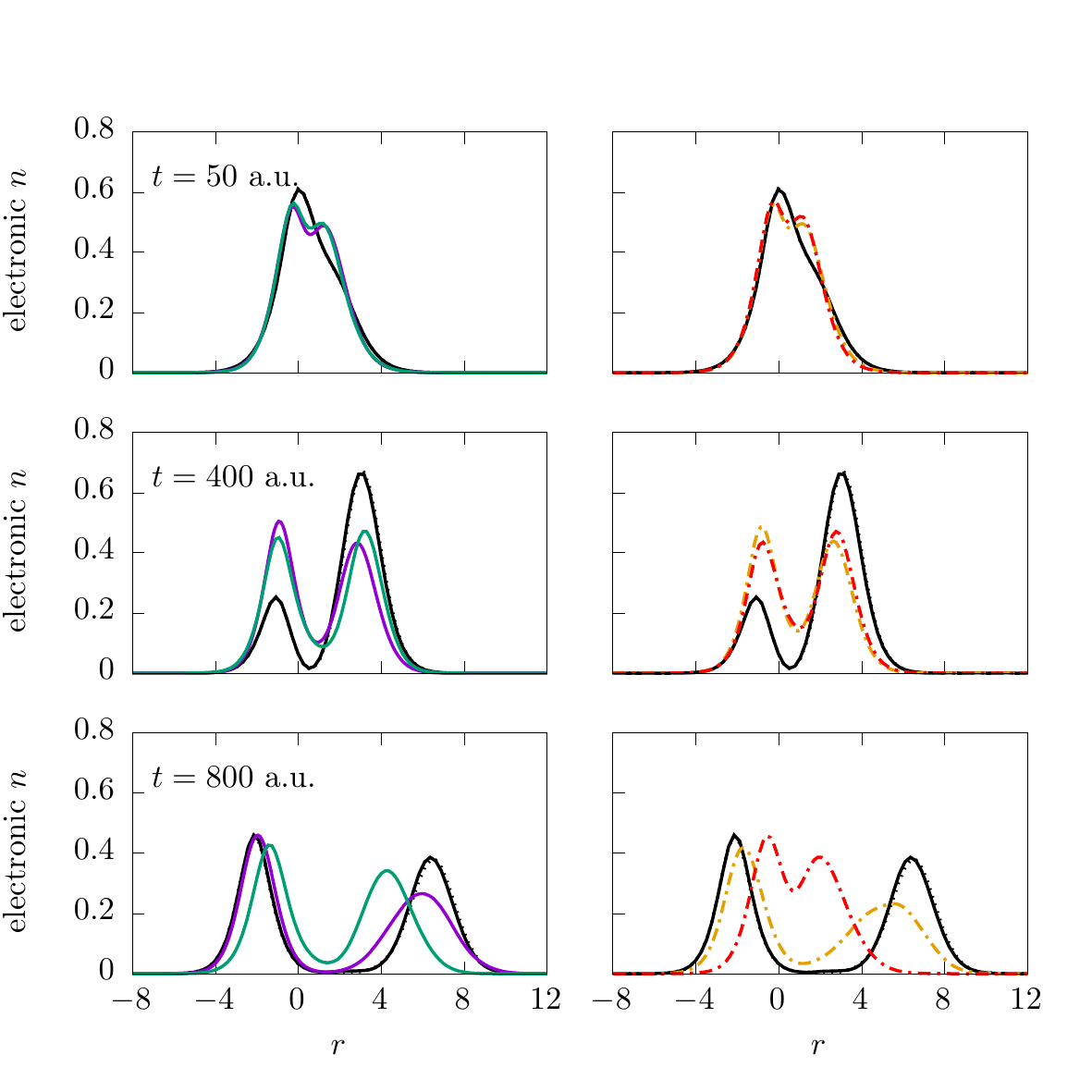}
\caption{Snapshots of electronic densities for different times. Left column: exact (black dotted), exact-Ehrenfest (black solid), LR ALDA (violet solid), RT ALDA (green solid). Right column: exact (black dotted), exact-Ehrenfest (black solid), LR AEXX (orange dash dotted), RT AEXX (red dash dotted). }
\label{fig:snaps}
\end{figure}

The fundamental reason that the LR-Ehrenfest dynamics is better than the RT can be understood from the earlier argument considering the domains of the xc functionals involved in the two types of calculations. 
Our results imply that when using TDDFT with adiabatic functionals, LR-Ehrenfest calculations are preferable to RT ones. 

\begin{figure}
\includegraphics[width=1.0\columnwidth]{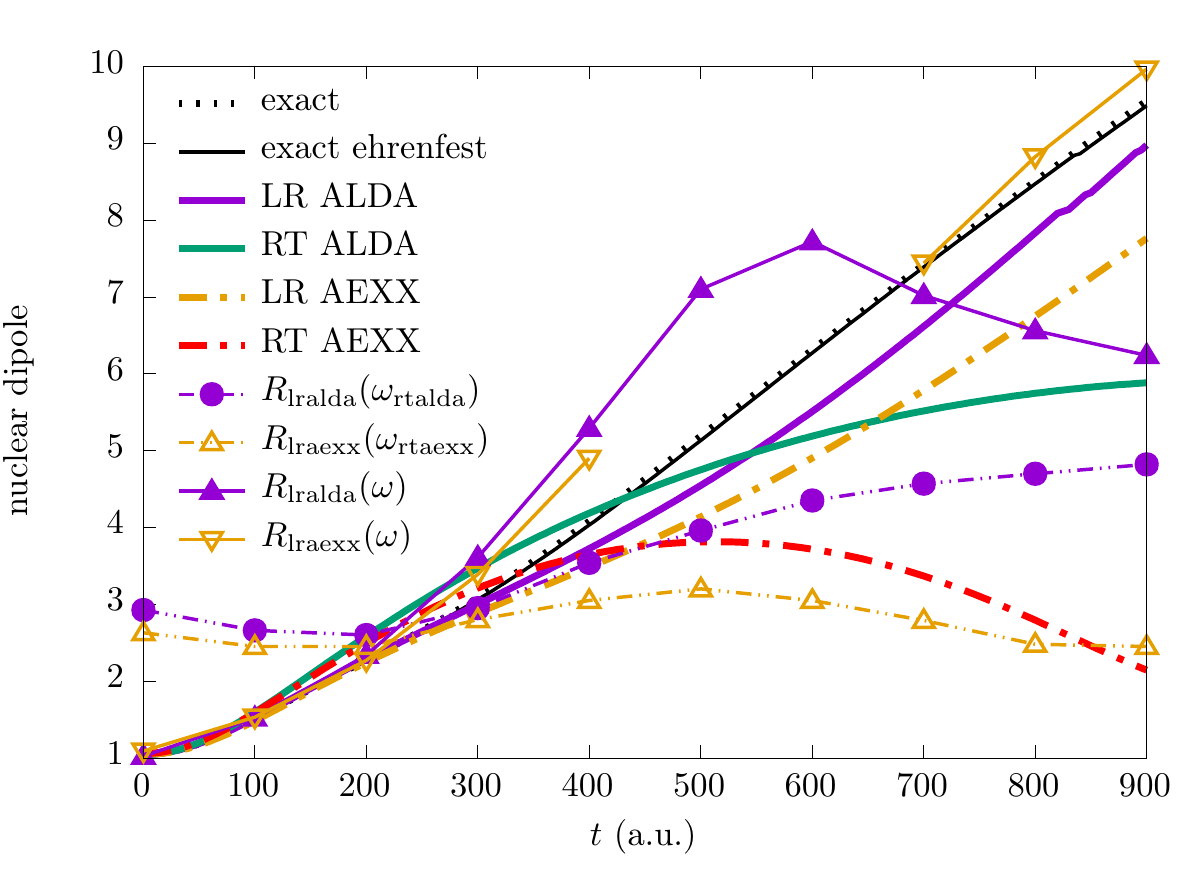}
 \caption{Deduced time-dependent nuclear  dipole:  $R_{\rm lralda}(\omega)$ , $R_{\rm lraexx}(\omega)$, $R_{\rm lralda}(\omega_{\rm tdalda})$,  $R_{\rm lraexx}(\omega_{\rm tdaexx})$ compared with exact, exact Ehrenfest, RT ALDA, RT AEXX, LR ALDA, LR AEXX with colors as indicated in the legend.}
 \label{fig:implieddip}
\end{figure}

As a last, dramatic, illustration of this, consider the application to femtosecond pump-probe spectroscopy set-ups, where the prospect of probing nuclear motion through electronic spectra has been raised~\cite{PSMCHSCTBWW17,ABPSCPL17,TBWLCPNLPG19}. The idea is that by comparing the measured time-resolved absorption spectrum of a molecule with the calculated spectrum at different nuclear configurations one can track the nuclear geometries as a function of time. If LR TDDFT is used to calculate the spectrum, then Fig.~\ref{fig:implieddip} plots the deduced nuclear separation $R_{\rm lr tddft}(\omega)$ where $\omega$ is the time-resolved excitation energy of the molecule at time $t$. At each time,  $\omega$ is computed from the field-free linear response of the molecule. We see that especially $R_{\rm lraexx}(\omega)$ does a reasonably good job except where at nuclear configurations whose exact frequencies go below the AEXX minimum frequency, while $R_{\rm lralda}(\omega)$ has some error especially at later times as expected due to the difficulties getting long-range charge-transfer excitations correct, but does a good job at earlier times. On the other hand, if 
 RT-Ehrenfest was used to simulate the  ``molecular movie", both $R_{\rm lraexx}(\omega_{\rm rtaexx})$ and $R_{\rm lralda}(\omega_{\rm rtalda})$ are qualitatively wrong from the very start. The problem with RT-Ehrenfest is related to the spurious peak shifts that have been observed in TDDFT when the system has been driven far from a ground-state~\cite{FLSM15,LFM16,HTPI14,PI16}.

In summary,  LR-Ehrenfest  calculations perform significantly better than RT-Ehrenfest simulations when using adiabatic xc functional approximations, because the xc functional is evaluated on a domain that  is closer to that from which they were derived. This affects qualitative predictions of coupled electron-nuclear dynamics in a number of applications, including time-resolved spectroscopy to probe ionic motion. Thus, LR-Ehrenfest calculations should be chosen over RT ones wherever possible. 
Still, we note two disadvantages of LR-Ehrenfest. The first is that non-adiabatic couplings between excited states should be obtained via quadratic response theory, where the adiabatic approximation can yield unphysical divergences~\cite{LL14,LSL14,ZH15,OBFS15}. Secondly, when external fields are present, LR-Ehrenfest is not as easily generalizable as RT-Ehrenfest. 
In a broader view, we expect some of the issues we highlighted here to be relevant in any self-consistent approach to excited-state dynamics, for example in time-dependent Hartree-Fock and orbital-dependent functional approaches. In these situations also, a LR-Ehrenfest calculation should be superior.  
\acknowledgements
{Financial support from the National Science Foundation Award CHE-1940333 (NTM)  and from the Department
of Energy, Office of Basic Energy Sciences, Division of Chemical
Sciences, Geosciences and Biosciences under Award No. DESC0020044 (LL) are gratefully acknowledged. }
\bibliography{./ref.bib}

\end{document}